\documentclass[%
 reprint,
superscriptaddress,
showpacs,
 amsmath,amssymb,
 aps,
]{revtex4-1}

\usepackage{graphicx}
\usepackage{color}
\usepackage{verbatim}
\usepackage{subfigure}

\begin{document}

\preprint{APS/123-QED}

\title{Photonic crystal nanobeam cavities with optical resonances around 800\,nm}

\author{I. Saber}
\affiliation{Centre de Nanosciences et de Nanotechnologies CNRS/Universit\'e Paris-Saclay, 91120 Palaiseau, France}
\author{R. Boddeda}
\affiliation{Laboratoire Kastler Brossel, Sorbonne Universit\'e, CNRS, ENS-PSL Research University, Coll\`ege de France, 4 place Jussieu, 75252 Paris, France}
\author{F. Raineri}
\affiliation{Centre de Nanosciences et de Nanotechnologies CNRS/Universit\'e Paris-Saclay, 91120 Palaiseau, France}
\affiliation{Universit\'e Paris Diderot, Sorbonne Paris Cit\'e, 75013 Paris, France}
\author{D. Sanchez}
\affiliation{Centre de Nanosciences et de Nanotechnologies CNRS/Universit\'e Paris-Saclay, 91120 Palaiseau, France}
\author{G. Beaudoin}
\affiliation{Centre de Nanosciences et de Nanotechnologies CNRS/Universit\'e Paris-Saclay, 91120 Palaiseau, France}
\author{I. Sagnes}
\affiliation{Centre de Nanosciences et de Nanotechnologies CNRS/Universit\'e Paris-Saclay, 91120 Palaiseau, France}
\author{Q. Glorieux}
\affiliation{Laboratoire Kastler Brossel, Sorbonne Universit\'e, CNRS, ENS-PSL Research University, Coll\`ege de France, 4 place Jussieu, 75252 Paris, France}
\author{A. Bramati}
\affiliation{Laboratoire Kastler Brossel, Sorbonne Universit\'e, CNRS, ENS-PSL Research University, Coll\`ege de France, 4 place Jussieu, 75252 Paris, France}
\author{J.~A.~Levenson}
\affiliation{Centre de Nanosciences et de Nanotechnologies CNRS/Universit\'e Paris-Saclay, 91120 Palaiseau, France}
\author{K.~Bencheikh}\email{kamel.bencheikh@c2n.upsaclay.fr}
\affiliation{Centre de Nanosciences et de Nanotechnologies CNRS/Universit\'e Paris-Saclay, 91120 Palaiseau, France}

\begin{abstract}
We report on the design and the fabrication of 1D photonic crystal (PhC) nanobeam cavities with optical resonances around 800\,nm, compatible with Rubidium, Cesium or Argon atomic transitions. The cavities are made of Indium Gallium Phosphide (InGaP) material, a III-V semi-conductor compound which has a large index of refraction ($n \simeq 3.3$) favoring strong optical confinement and small mode volumes. Nanobeam cavities with inline and side coupling have been designed and fabricated, and quality factors up to $2\times 10^4$ have been measured.
\end{abstract}

\maketitle

\section{Introduction}

Cavity Quantum Electrodynamics (CQED) is based on the strong interaction between single atoms and photons\cite{Kimble1998,Walther2006}. The ability to reach this regime mostly relies in the possibility of confining the electromagnetic field in optical resonators with long-lived photons. Over the last years, impressive results have been achieved in terms of the figure of merit $Q/V$, by reducing the cavity dimensions and by improving the reflectivity of the constitutive mirrors. To obtain high $Q$ factors with small mode volumes is one of the primary assets of Photonic Crystal (PhC) cavities. However, due to both technologic and application reasons, PhC are mostly developed at telecom wavelengths. Shorter wavelengths, suitable for example for strong coupling with cold atoms, are by far much more challenging in terms of technological accuracy \cite{Rousseau2017}.  

Photonic crystal cavities are generally obtained by creating an optical defect in a structure with a periodic change in the refractive index distribution\,\cite{Yablonovitch1987,John1987,joannopoulos2011photonic}. When the optical frequency of the electromagnetic field is in the band-gap of the PhC, light is trapped in the defect region \cite{Foresi1997, Joannopoulos1997}. Various cavities based on this principle have been designed and fabricated. Popular designs are based on 2D PhC, where one or several holes are removed from the otherwise periodic lattice\,\cite{Painter1819,  Asano2006, Galli2014} or  2D PhC waveguides, where a full row of holes in the periodic lattice is removed\,\cite{Kuramochi2006, Asano2017,Noda2014}. In the latter, the cavity is obtained by engineering the period, largest in its center and decreasing  when moving away. The propagating mode of the waveguide at the center becomes evanescent at the edges and is thus confined, forming a resonant mode. Cavities are also obtained in 1D PhC, usually made in a ridge waveguide in which two series of air-holes are etched\,\cite{Ahn2010, Halioua2010, Quan2010, Quan2011,Crosnier2016}. The two series are usually separated by a distance $L$ forming the cavity, where light is trapped. Such cavities, usually referred to as nanobeam cavities, benefit from the smallest mode volume and are thus of primary interest for CQED. 

In this paper, we report on the design and fabrication of high $Q$ PhC cavities with small mode volumes with optical resonances in the low infrared region ($\sim\,$800\,nm) which make them  ideal to interact  with alkali atoms such as Rubidium and Cesium and with Argon atoms. Our design is based on a typical 1D PhC cavity. However for a PhC with a fixed period, the achievable $Q$ factors are quite limited. Indeed, the theory based on Fourier space analysis shows that the $Q$ factors of the PhC cavities are limited by the out-of-plane scattering \cite{Vuckovic2002, Srinivasan2002}. To minimize the scattered power, one needs to attenuate the Fourier components in $k$ space below the light cone. This is obtained when the mode field has a Gaussian profile in real space \cite{Akahane2003, Akahane2005, Asano2006}. In practice, such mode profile can be achieved by engineering the PhC  \cite{Bazin2014,Quan2011} throughout the modification of one or several of its parameters, namely the period, the hole radius, the width, or the thickness. In our design, we modulate only the PhC period along the propagation direction \cite{Bazin2014}. In order to reduce the cavity mode volume while minimizing the losses, our nanobeam cavities are designed in such a way that their length $L$ is  a single period of the PhC, as demonstrated by Q. Quan \emph{et. al.} in Ref.\,\cite{Quan2010}. The coupling into the nanocavity is an important issue to take into consideration when developing high-$Q$ cavities. Indeed, as the unavoidable coupling opens an extra lossy channel, it will strongly influence the measured cavity $Q$ factor. The coupling will also set the transmission amplitude to and from the cavity. The most trivial coupling way is the inline coupling, where the ridge waveguide in which the nanobeam cavity  is made also serves as the input and output channels. Recently, it has been shown that in such design the transmission of the fundamental mode is rather low\cite{Afzal2018}. This has been confirmed by our measurements and it is due to the wavevector mismatch between the  resonant mode in the cavity and the propagating mode in the waveguide. In order to circumvent this transmission problem, we have also designed nanobeam cavities with side coupling. Ridge waveguides in the vicinity  and on both sides of  the nanobeam cavities are fabricated to couple in the light from a conventional laser source and to couple out the light from the cavities. To launch or extract light into the coupling waveguides, Bragg gratings are fabricated at their extremities as they are routinely used at telecom wavelengths\,\cite{Baets2006}. When properly designed, they could show extremely high coupling efficiencies\,\cite{Peucheret2013, Andreani2015}. However in this paper, we payed little attention to their design, focusing all our efforts on the development of the nanobeam cavities.

In the next section, we will first describe in detail the design to improve the $Q$ factors of 1D PhC nanobeam cavities and present their fabrication process. The optical characterizations will be described in the last section, before the conclusion.

\section{Design and fabrication}
\label{sec:examples}
In order to achieve nanobeam cavities with optical resonances close to 800\,nm and small modal volumes, one needs to use a high-index compound transparent in the near infrared band. We have chosen Indium Gallium Phosphide (InGaP) because of its high index of refraction, $n = 3.32$ at 800\,nm, and its electronic band-gap of 1.85\,eV which makes it transparent for wavelengths higher than 674\,nm. The InGaP nanobeam is 150\,nm thick, about $\lambda/(2n)$, ensuring single mode operation, and 300\,nm wide. The 1D PhC is obtained by etching 120\,nm-diameter circular air-holes in the InGaP. For the $Q$ factor optimization, the periods of the PhC were varied from 180\,nm at the center to 215\,nm at the edge of the nanobeam cavity,  according to an algorithm that allows the mode to have a Gaussian shape, following the apodization technique described in Ref.\,\cite{Bazin2014, Akahane2003}. The algorithm gives the period as a function of the position in the PhC mirror. The out-of-plane losses are thus reduced by minimizing the scattered power given by the integral of the spatial Fourier components of the field within the light cone. Following the analysis in Ref.\,\cite{Bazin2014}, the first period is chosen in such a way that the target resonance frequency ($f_0 = 375\,$THz, $\lambda_0 = 800$\,nm) of the cavity coincides with the cut-off frequency of the dielectric mode of the PhC at $k =0.5$ ($2 \pi/a$). The field has an evanescent attenuation in the band-gap of the 1D PhC crystal, with an enveloppe decaying as $e^{-q x}$ where $q$ is the imaginary part of the wavevector.

\begin{figure}[ht]
\centering
\includegraphics[width=0.8\linewidth]{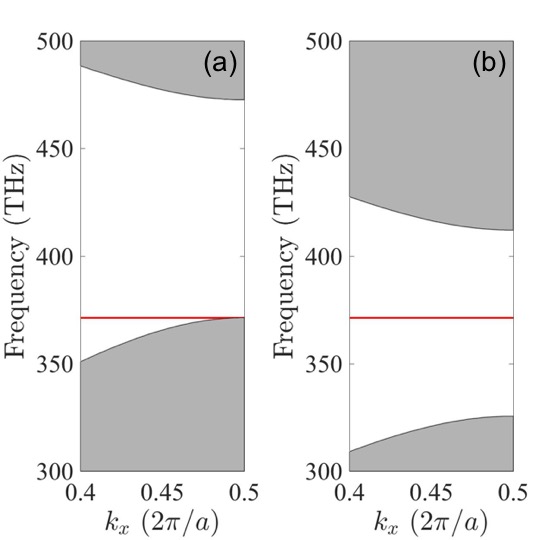}
\caption{Band diagram for a 1D PhC in a InGaP ridge waveguide having a width of 300\,nm and a thickness of 150\,nm. The periodicity of the PhC is 180\,nm in (a) and 215\,nm in (b). The radius of the air-holes is 60\,nm. The red line shows the target resonance frequency at 375\,THz corresponding to a wavelength of 800\,nm. The gray areas are the dielectric (bottom) and air (top) bands.}
\label{fig:db} 
\end{figure}

\begin{figure}[h]
\centering
\includegraphics[width=1\linewidth]{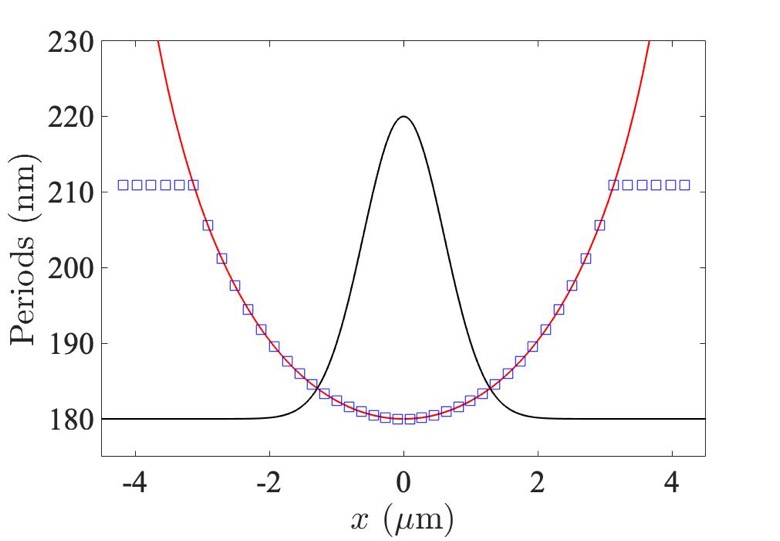}
\caption{\label{fig:periodsvsx} Evolution of the period as a function of the mirror lengths. $x=0$ corresponds to the center of the nanocavity. The red line is the theoretical behavior and the blue squares are the real periods as a function of the real air holes positions. The black line is the Gaussian mode profile with a FWHM of 1\,$\mu$m.}
\end{figure}

At the cut-off frequency, attenuation constant $q$ is 0. Figure\,\ref{fig:db}(a) shows the band diagram for 1D  PhC with a period $a=180$\,nm.  The red line indicates the target resonance frequency $f_0$ and the gray regions are the dielectric (bottom) and air (top) bands of the PhC. The strongest value of $q$ is obtained for a PhC with a period $a=215$\,nm, where $f_0$ lies in the middle of the band-gap as shown in Fig.\,\ref{fig:db}(b).  In order to have a field with a Gaussian profile, the attenuation $q$ needs to be varied linearly along the $x$ direction of the PhC, $q = B x$, so that the exponential decay $e^{-q x}$ becomes a Gaussian attenuation $e^{-B x^2}$. 
Our algorithm allows  to determine the evolution of the periods of the PhC from 180\,nm to 215\,nm along the $x$ direction to obtain a Gaussian field variation for a given full width at half maxima of the optical mode field, which depends on $B$, the only free parameter. The continuous red line in Figure \,\ref{fig:periodsvsx} shows the evolution of the period as a function of the position in the PhC mirror surrounding the cavity when the FWHM of the mode is chosen to be 1\,$\mu$m. The squares are the actual positions of the air-holes centers following the theoretical line. Additional air-holes, 5 in our case, with the maximal period are added at the end of the PhC mirrors in order to further increase the $Q$ factor. The black line indicates the intensity profile of the Gaussian mode.

\begin{figure}[htb]
\centering
\includegraphics[width=1\linewidth]{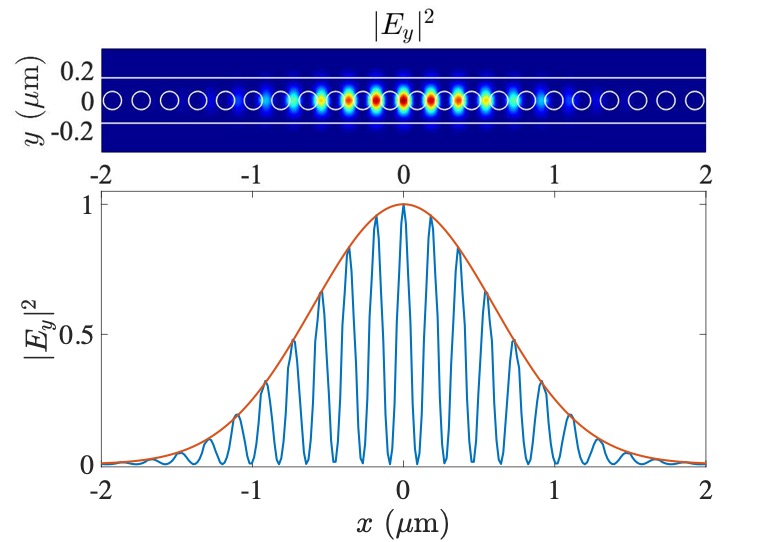}
\caption{\label{fig:modeprofile} 2D profile of the fundamental mode at 811\,nm in the $xy$ plan (top) and in the $x$ direction for $y=0$ (bottom, blue line). The red line represents a  Gaussian envelope of the mode with a FWHM of 1\,$\mu$m in intensity.}
\end{figure}

The optical properties of the designed InGaP nanobeam cavity are evaluated using three-dimensional Finite-Difference Time-Domain (3D FDTD) simulations implemented with LUMERICAL FDTD. We obtain a fundamental optical resonance at 811\,nm, close to the target value, with a quality factor $Q =1.5\times 10^7$ and a mode volume of 0.01\,$\mu$m$^3$ which is about $0.64\,(\lambda/n)^3$. The second and third-order modes are at 823\,nm and 836\,nm, and have respectively $Q$-factors of $3.6\times 10^6$ and $2\times 10^6$ with mode volumes $0.81\,(\lambda/n)^3$ and $0.91\,(\lambda/n)^3$. Figure\,\ref{fig:modeprofile} shows the fundamental mode profile in the 2D $xy$ plane of the cavity. The mode is mainly located between the air-holes in the dielectric as expected since we have fixed the target resonance frequency close to the dielectric band. The  field intensity is maximum at the middle of the cavity where the period is $a = 180\,$nm. Moreover, it has a Gaussian envelope as indicated by the red line.

\begin{figure*}
    \centering
    \includegraphics[width = 1\linewidth]{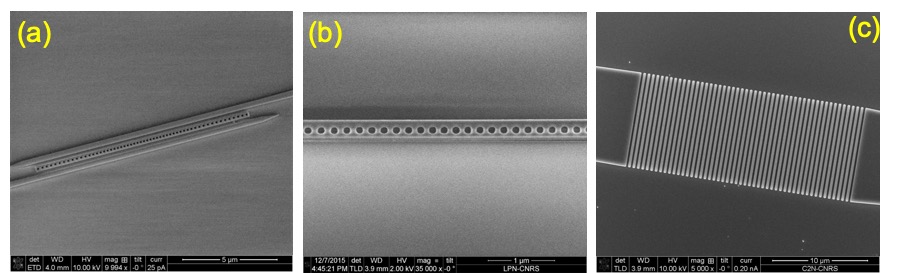}
    \caption{Scanning Electron Microscope (SEM) image of (a) a nanocavity with two adjacent waveguides, (b) a nanocavity in the coupling waveguide, and (c) a typical Bragg grating for coupling from free space into the different nanocavities.}
    \label{fig:SEM}
\end{figure*}

The nanocavities are fabricated by first growing InGaP on a GaAs substrate using metal-organic chemical vapor deposition. The InGaP/GaAs is then bonded to a Silicon on insulator Si/SiO$_2$ substrate using benzocyclobutene (BCB) polymer layer. Finally the GaAs is removed using wet etching, obtaining thus the final 150\,nm-thick InGaP layer on an SOI substrate. Electron-beam lithography is used afterwards to draw the designed PhC structures on the InGaP covered with a thin layer of electron-sensitive resist. After removing the non-exposed resist, the InGaP is etched using Inductively Coupled Plasma Reactive Ion Etching (ICP-RIE). Figure \ref{fig:SEM} shows  Scanning Electron Microscope (SEM) images of typical nanobeam cavities we have fabricated: (a) with two adjacent waveguides intended for evanescent side coupling to the cavity; (b) with inline waveguide for direct coupling.  Figure \ref{fig:SEM} (c) represents a SEM image of a typical grating fabricated at the extremities of the coupling waveguides. The Bragg gratings are used to couple in and out light to the waveguide from free space, usually from an optical fiber. The periodicity of the Bragg grating is 470\,nm with a 50\,\% duty cycle, so that the diffraction angle is $\sim 10^{\circ}$ off the normal of the grating around $\lambda \approx 800$\,nm  in order to avoid back coupling into the fiber due to the second order diffraction. The Bragg gratings are fabricated simultaneously with the nanobeam cavity by fully etching the InGaP on half the period. The gratings are 10\,$\mu$m wide and 15\,$\mu$m long. For a lossless propagation from the grating couplers to the nanobeam cavity, a 300-$\mu$m long optical taper is used between the 10-$\mu$m large gratings and the narrow ($\sim$ 300\,nm) width coupling waveguides \cite{Fu2014}.

During the fabrication process, we have varied different parameters  to study their effect on the optical properties of PhC. In the case of the inline nanobeam cavities, we  varied the periods of the PhC to tune the wavelengths of the optical resonances. More specifically, we change the central period $a$ from 176\,nm to 188\,nm while keeping the maximal period at the edges at 215\,nm. When increasing the period, the PhC band-gap is shifted towards lower energies. As a consequence, the resonant wavelengths, and more particularly the fundamental mode, are tuned to the red. For the nanobeam cavities with the side coupling strategy, we varied the width $L$ of the coupling waveguides and the distance $d$ between the cavity and the coupling waveguides. The parameters of the cavity, such as the central period of 180\,nm and the maximal period of 215\,nm, are kept constant. By construction, the nanobeam cavity and the coupling waveguides have identical thickness. When varying the width $L$, the wavevector $k_w$ of the mode propagating in the coupling waveguides is tuned. The largest coupling to the cavity is obtained when $k_w$ matches with the wavevector of the fundamental resonant mode\,\cite{Crosnier2016}, i.e. $k_0  = \pi/a$. As the period of the PhC mirrors of the cavity is not constant, $a$ is determined by taking the Fourier transform of the mode represented in Fig.\,\ref{fig:modeprofile} (bottom sketch), obtaining $a =182.6$\,nm. $k_w =  \pi/a$ is achieved for a waveguide width $L = 220$\,nm.  The distance $d$ between the nanobeam cavity and the coupling waveguide will modify the integral overlap between the mode field of the cavity and the mode field of the coupling waveguide. Both $L$ and $d$ will impact the coupling strength which can be quantified by an effective quality factor $Q_c$.  The quality factor that will be measured experimentally is then given by $1/Q = 1/Q_0+1/Q_c$, where $Q_0$ is the intrinsic quality factor of the nanobeam cavity.

\begin{figure}[h]
    \centering
    \includegraphics[width = 1\linewidth]{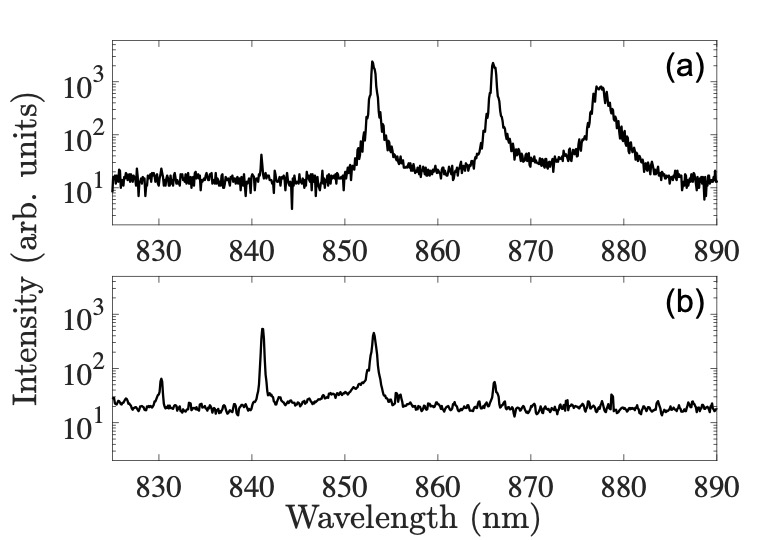}
    \caption{Transmission spectra for an inline coupling nanobeam cavity. The light is collected with microscope objective above the output Bragg grating (a), or above  the nanobeam cavity itself (b). The fundamental mode at 830\,nm is not observed when light is collected at the output grating.}
    \label{fig:TRANSCAV}
\end{figure}

\section{Experimental observations}
\label{experiment}

To investigate the fabricated nanobeam cavities, we used two laser sources. A pulsed Ti:Sapphire laser delivering  20-nm-width pulses at a repetition rate of 80\,MHz and a continuous-wave (cw) diode laser source tunable between 790\,nm and 850\,nm. The pulsed source is used to identify rapidly  the wavelengths of the resonances by analyzing the spectrum of the pulses out of the nanobeam cavities using a 40\,pm resolution spectrometer (Ocean Optics HR2000). For the measurement of the resonance linewidths and hence the associated $Q$ factors, we used the CW diode laser with a resolution of 1\,pm. A wavelength-meter (\AA ngstrom WS6-600) with a measurement resolution of 20\,MHz is used to read accurately the wavelength. To couple the light into the nanocavities we used a cleaved single mode fiber (Thorlabs 780HP), set at an incidence angle of about $10^{\circ}$ in front of the Bragg grating. The laser is coupled to the nanobeam cavity after diffracting from the Bragg grating and propagating along the optical taper and in the coupling waveguides. The transmitted light is collected using a long working distance microscope objective (Mitutoyo M Plan NIR 10$\times$) positioned  above the output Bragg grating. We estimate the transmission to be about 0.04\,\%. It is inferred from the measurement of the power incident on the Bragg grating and the power collected after the microscope objective. The collected light is then coupled to a fiber and sent either to the spectrometer or to a high sensitivity Si photodiode. 

\begin{figure}
\centering
\includegraphics[width=1\linewidth]{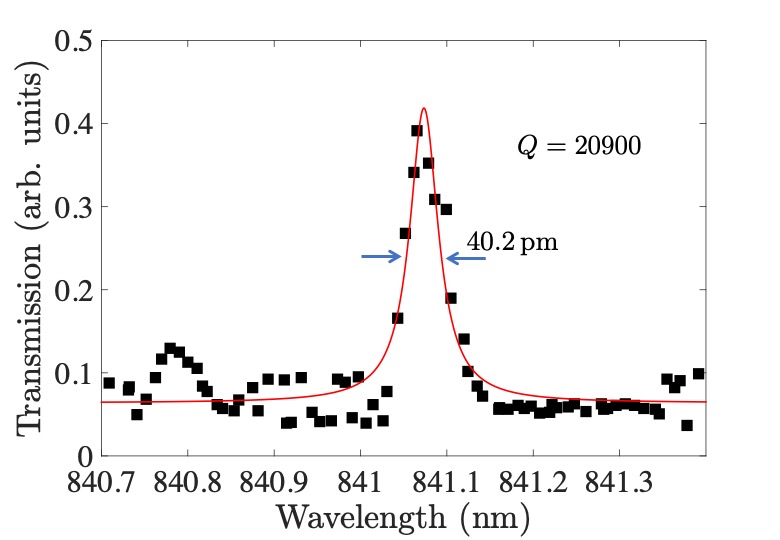}
\caption{\label{fig:res840nm} Transmission of the second resonant mode of the nanbeam cavity (see Fig.5). From the Lorentzien ajustement  (red line) we can extract the resonant wavelength $\lambda = 841.07$\,nm, the FWHM $\Delta \lambda = 40.2$\,pm, which corresponds to a quality factor $Q = 20900$. }
\end{figure}

When working with the inline nanobeam cavities, the transmission of the fundamental mode is weak and hard to observe. Figure\,\ref{fig:TRANSCAV} shows a typical spectrum measured with the spectrometer. In the spectrum of Fig.\,\ref{fig:TRANSCAV} (a), the light was collected at the output Bragg grating with the microscope objective. For this particular nanobeam cavity the fundamental mode is at 830\,nm. However, due to the weakness of the coupling it is not observable. The first observable resonance is associated to the second-order mode at 841\,nm as further shown in the spectrum of Fig.\,\ref{fig:TRANSCAV} (b) obtained by collecting directly the light leaking above the cavity.
Note that in this case the fundamental mode although weak, is clearly observable. The weak transmission of the fundamental mode is due to the wavevector mismatch between the resonant mode and the mode propagating in the waveguide holding the nanobeam cavity which is 300\,nm width. Furthermore, the spatial size of the fundamental mode along the propagation direction is very small (1\,$\mu$m FWHM) compared to the dimension of the nanobeam cavity as shown in Fig.\,\ref{fig:periodsvsx}.  The resonant mode is thus strongly confined and hard to extract from the nanocavity. We did not succeed to measure the $Q$ factor of the fundamental mode because of the weakness of signal, even when collecting it above the cavity. We have nevertheless measured the $Q$ factor of the second-order mode at 841\,nm using the cw laser. Figure\,\ref{fig:res840nm} shows the obtained spectrum from which we can deduce a quality factor $Q= 20900$. According to the numerical FDTD simulations, the $Q$ factor of the second-order mode is one order of magnitude smaller than that of the fundamental mode, we can thus expect a larger $Q$ factor at 830\,nm. However the measured $Q$ factors are still below  the numerical simulations by three to four orders of magnitude. This is a common trend when working with such very small dimensions PhC systems in which the propagating short-wavelength light becomes very sensitive to the surface imperfections. Indeed the cavity mode is not tightly confined in the so small cross-section of the nanobeam cavities. It is as well delocalized in the InGaP/SiO$_2$ interface presenting surface roughness causing extra losses.

\begin{figure}[h]
    \centering
    \includegraphics[width = 1\linewidth]{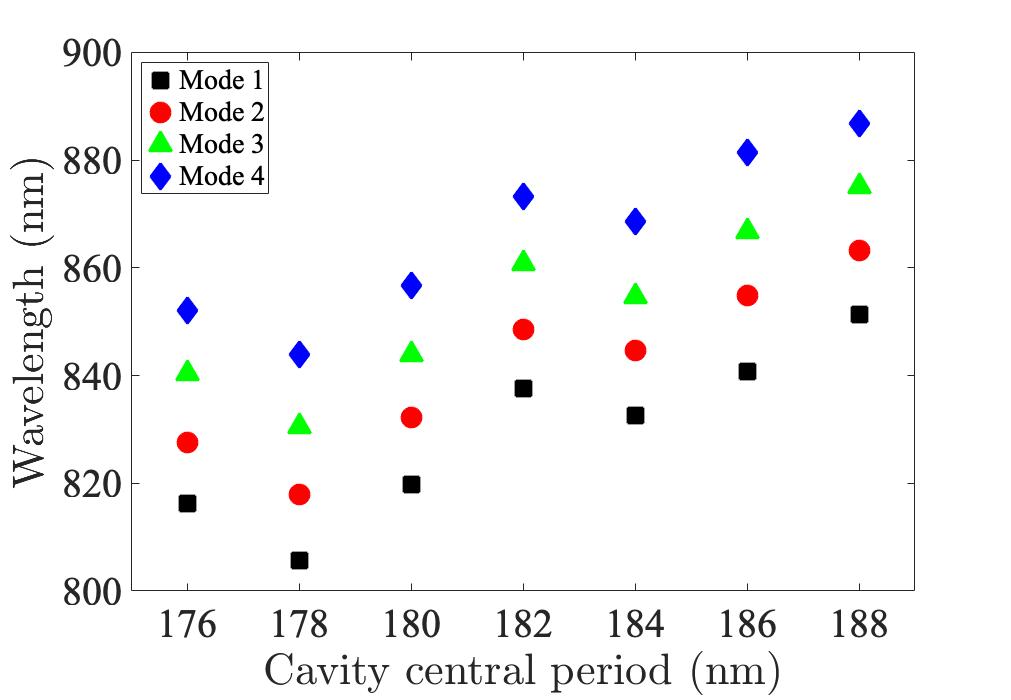}
    \caption{The first four resonant modes measured for an inline coupling nanobeam cavity when its central period is varied.}
    \label{fig:modesvsperiod}
\end{figure}

\begin{figure}[h]
\centering
\includegraphics[width=1\linewidth]{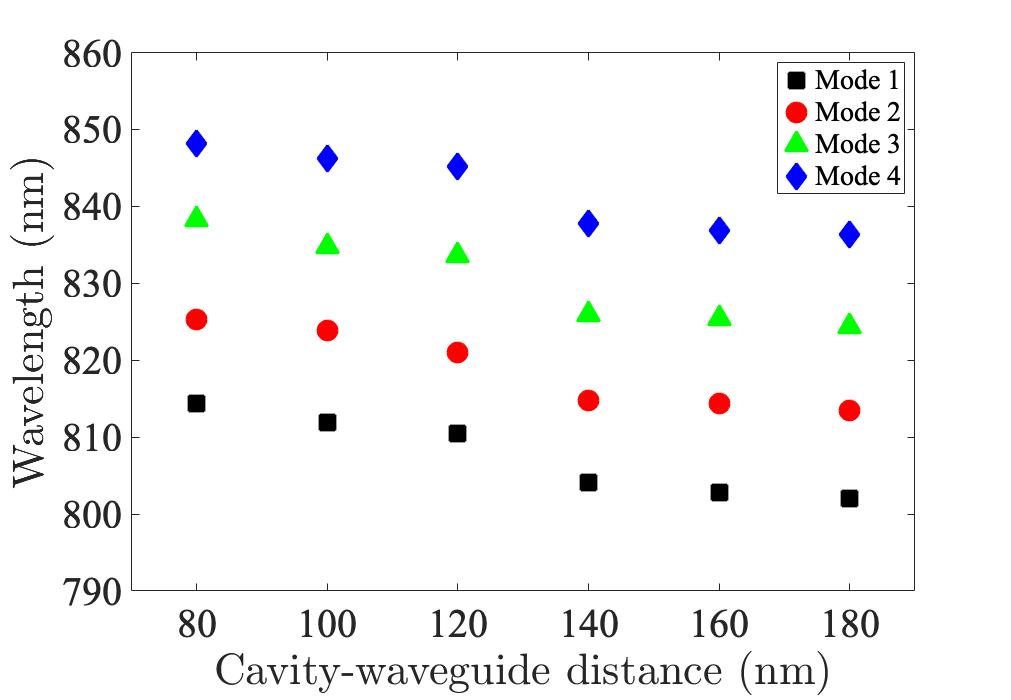}
\caption{\label{fig:wvlvsd} The first four resonant modes for side coupled nanobeam cavity as a function of the cavity-waveguide distance. The width of the coupling waveguide is $L = 250$\,nm.}
\end{figure}

\begin{figure}
\centering
\includegraphics[width=1\linewidth]{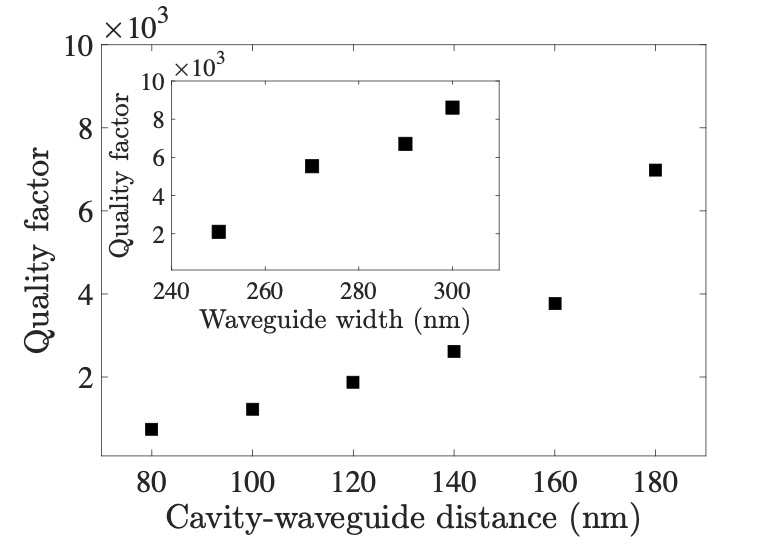}
\caption{\label{fig:QvsdandL} Evolution of the $Q$ factor as a function of the cavity-waveguide distance $L$ for a fixed waveguide width $d = 250$\,nm. Inset:  $Q$ factor as a function of the waveguide width $d$ for a fixed distance $L=120$\,nm between the nanobeam cavity and the coupling waveguide.}
\end{figure}

As expected, when increasing the central period of the inline nanobeam cavity, the wavelengths of the resonant modes are tuned, shifting to the infrared. This is indeed represented in Fig.\,\ref{fig:modesvsperiod} for the first four modes we have measured using the pulsed source and the spectrometer. The depicted resonances are obtained by measuring mainly the transmitted light collected at the output Bragg grating.  Additional spectra  obtained by measuring the cavity leakage were however necessary to confirm the wavelength of the fundamental mode. Similar measurements are done on the nanobeam cavities with the side coupling strategy. Figure\,\ref{fig:wvlvsd} shows the first four modes measured using the pulsed laser source as a function of the cavity-waveguide distance. In the case of the side-coupled cavities, all the modes, and especially the fundamental one, are very well resolved. 
The coupling waveguides have a width of 250\,nm. As the coupling waveguide gets closer to the cavity, the wavelengths of the modes shift to the infrared. Indeed  the nanobeam cavity's effective index of refraction is altered, modifying the optical properties of the nanobeam cavity and hence its resonant wavelength. As the distance is increased, the coupling between the nanobeam cavity and the waveguide gets weaker, with negligible effects on the resonant modes. When the coupling waveguides are close to the nanobeam cavity, the transmission is  higher. However the resonances become broader due to the large leakage which reduces the photons lifetime in the cavity. This is shown in Fig.\,\ref{fig:QvsdandL} where the $Q$ factor of the fundamental mode is plotted as a function of the distance $d$ for a waveguide width $=250$\,nm. The quality factors are deduced by measuring the FWHM of the resonances using the cw laser. We show also in the inset that the $Q$ factor goes down as the width of the coupling waveguide is reduced. The quality factor should go down until $d \sim 200$\,nm is reached, width for which perfect phase matching is achieved between the resonant mode and the guided mode and  hence optimum coupling. The best $Q$ factor is below 10000 for the side coupled nanobeam cavities, a factor of two smaller than the $Q$ factor of the inline nanobeam cavity. This is due to technological imperfections and the presence of extra coupling channels to the side waveguides. 

\section{Conclusion}
In summary, we have proposed high-$Q$ PhC nanobeam cavities with optical resonances at near infrared wavelengths around 800\,nm, ideally suited for performing cavity QED-like measurements with alkali atoms.We have fabricated cavities using side and inline coupling. The best measured $Q$ factor is  about 20000. Though it is three orders of magnitude smaller than the predicted value, it is one of the highest $Q$ factor measured at this short wavelength. We have shown that coarse tuning of the resonances can be easily achieved by a small change of the PhC parameters. Fine tuning can also be achieved by controlling the nanobeam cavities temperature.\\

\section*{Funding} 
Public grants CONDOR and ICQOQS overseen by the French National Research Agency (ANR) as part of the ``Investissements d'Avenir" program (reference: ANR-10-LABX-0035, Labex NanoSaclay); French ANR grant UNIQ DS078; The French RENATCH Network and the French ANR grant UNIQ DS078.\\

\section*{Acknowledgment} 
The authors gratefully acknowledge Alejandro Giacomotti for helpful discussions.


\begin{thebibliography}{99}

\bibitem{Kimble1998} H. J. Kimble, Phys. Scr. \textbf{1998}, 127 (1998).

\bibitem{Walther2006} H. Walther, B. T. H.Varcoe, B.-G. Englert, and T. Becker, Rep. Prog. in Phys. \textbf{69},1325 (2006).  
  
\bibitem{Rousseau2017}  I. Rousseau, I. S\'anchez-Arribas, K. Shojiki, J.-F. Carlin, R. Butt\'e, and N. Grandjean, Phys. Rev. B \textbf{95}, 125313 (2017).

\bibitem{Yablonovitch1987} E. Yablonovitch, Phys. Rev. Lett. \textbf{58}, 2059 (1987).

\bibitem{John1987} S. John, Phys. Rev. Lett. \textbf{58}, 2486 (1987).

\bibitem{joannopoulos2011photonic} J. Joannopoulos, S. Johnson,  J. Winn, and R. Meade, \textit{Photonic Crystals: Molding the Flow of Light}, Second Edition (Princeton University Press, 2011).

\bibitem{Foresi1997} J. S. Foresi, P. R. Villeneuve, J. Ferrera, E. R. Thoen, G. Steinmeyer, S. Fan, J. D. Joannopoulos, L. C. Kimerling, H. I. Smith, and E. P. Ippen, Nature \textbf{390}, 143 (1997).
 
\bibitem{Joannopoulos1997} J. D. Joannopoulos, P. R. Villeneuve, and S. Fan, Nature \textbf{386}, 143 (1997).
  
\bibitem{Painter1819} O. Painter, R. K. Lee, A. Scherer, A. Yariv, J. D. O'Brien, P. D. Dapkus, and I. Kim, Science \textbf{284}, 1819 (1999).

\bibitem{Asano2006} T. Asano, B. Song, Y. Akahane, and S. Noda, IEEE J. Sel. Top. Quantum Electron. \textbf{12}, 1123 (2006).  
  
\bibitem{Galli2014} Y. Lai, S. Pirotta, G. Urbinati, D. Gerace, M. Minkov, V. Savona, A. Badolato, and M. Galli, Appl. Phys. Lett. \textbf{104}, 241101 (2014).
  
\bibitem{Kuramochi2006} E. Kuramochi, M. Notomi, S. Mitsugi, A. Shinya, T. Tanabe, and T. Watanabe, Appl. Phys. Lett. \textbf{88}, 1 (2006).
  
\bibitem{Asano2017} T. Asano, Y. Ochi, Y. Takahashi, K. Kishimoto, and S. Noda, Opt. Express \textbf{25}, 1769 (2017).

\bibitem{Noda2014} H. Sekoguchi, Y. Takahashi, T. Asano, and S. Noda, Opt. Express \textbf{22}, 916 (2014).
  
  
\bibitem{Ahn2010} B.-H. Ahn, J.-H. Kang, M.-K. Kim, J.-H. Song, B. Min, K.-S. Kim, and Y.-H. Lee, Opt. Express \textbf{18}, 5654 (2010).
  
  
\bibitem{Halioua2010} Y. Halioua, A. Bazin, P. Monnier, T. J. Karle, I. Sagnes, G. Roelkens, D. Van Thourhout, F. Raineri, and R. Raj, J. Opt. Soc. Am. B \textbf{27}, 2146 (2010).
  
\bibitem{Quan2010} Q. Quan, P. B. Deotare, and M. Loncar, Appl. Phys. Lett. \textbf{96}, 203102 (2010).  
  
\bibitem{Quan2011} Q. Quan and M. Loncar, Opt. Express \textbf{19}, 18529 (2011).  
  
\bibitem{Crosnier2016} G. Crosnier, D. Sanchez, A. Bazin, P. Monnier, S. Bouchoule, R. Braive, G. Beaudoin, I. Sagnes, R. Raj, and F. Raineri, Opt. Lett. \textbf{41}, 579 (2016).
  
\bibitem{Vuckovic2002} J. Vuckovic, M. Loncar, H. Mabuchi, and A. Scherer, IEEE J. Quantum Elect. \textbf{38}, 850 (2002).
  
\bibitem{Srinivasan2002} K. Srinivasan and O. Painter, Opt. Express \textbf{10}, 670 (2002).  
  
\bibitem{Akahane2003} Y. Akahane, T. Asano, B.-S. Song, and S. Noda, Nature \textbf{425}, 944 (2003).  
  
\bibitem{Akahane2005} Y. Akahane, T. Asano, B.-S. Song, and S. Noda, Opt. Express \textbf{13}, 1202 (2005).  
  
\bibitem{Bazin2014} A. Bazin, R. Raj, and F. Raineri, J. Light. Technol. \textbf{32}, 952 (2014).
  
\bibitem{Afzal2018} F. O. Afzal, S. I. Halimi, and S. M. Weiss, J. Opt. Soc. Am. B \textbf{36}, 585 (2019).
  
\bibitem{Baets2006} D. Taillaert, F. VanLaere, M. Ayre, W. Bogaerts, D. Van Throurhout, P. Bienstman, and R. Baets, Jpn. J. Appl. Phys. \textbf{45}, 6071 (2006).  
  
\bibitem{Peucheret2013} Y. Ding, H. Ou, and C. Peucheret, Opt. Lett. \textbf{38}, 2732 (2013).  
  
\bibitem{Andreani2015} A. Bozzolla, L. Carroll, D. Gerace, I. Cristiani, and L. C. Andreani, Opt. Express \textbf{23}, 16289 (2015).
  
\bibitem{Fu2014} Y. Fu, T. Ye, W. Tang, and T. Chu, Photon. Res. \textbf{2}, A41 (2014).

\end{thebibliography}
\end{document}